\documentclass[sort&compress,final,1p]{elsarticle}
\usepackage{amsfonts}
\usepackage{amsmath}
\usepackage{amssymb}
\usepackage{graphicx}
\journal{Chemical Physics}
\begin{document}

\begin{frontmatter}

\title{An efficient method for quantum transport simulations in the time domain}

\author[BCCMS]{Y. Wang}
\author[BCCMS,HK] {C.-Y. Yam}
\author[BCCMS]{Th. Frauenheim}
\author[HK]{G.H. Chen}
\author[RATI]{T.A. Niehaus \corref{cor1}}
\cortext[cor1]{Corresponding author \ead{thomas.niehaus@physik.uni-regensburg.de}}

\address[BCCMS]{Bremen Center for Computational Materials Science,
Universt\"{a}t Bremen, Germany}
\address[HK]{Department of Chemistry, The University of Hong Kong,
Hong Kong, China}
\address[RATI]{University of Regensburg, 93040 Regensburg, Germany}

\begin{abstract}
An approximate method based on adiabatic time dependent density functional theory (TDDFT) is presented, that allows for the description of the electron dynamics in nanoscale junctions under arbitrary time dependent external potentials. In this scheme, the density matrix of the device region is propagated according to the Liouville-von Neumann equation. The semi-infinite leads give rise to  dissipative terms in the equation of motion which are calculated from first principles in the wide band limit. In contrast to earlier {\em ab-initio} implementations of this formalism, the Hamiltonian is here approximated  by a density expansion in the spirit of the density functional based tight-binding (DFTB) method without introducing empirical parameters. Results are presented for two prototypical molecular devices and compared to calculations at the full TDDFT level. The issue of non-existence of a steady state under certain conditions  is also briefly touched on.
\end{abstract}
\end{frontmatter}

\section{Introduction}
Molecular electronics -- the use of single molecules as functional
entities in electronic devices -- is often heralded as a replacement
for the conventional CMOS technology which faces physical limits in
further miniaturization. Even if the industrial application of  this
technology seems to be out of reach in the moment, research on the
electronic transport through individual molecules has already revealed
a variety of interesting quantum phenomena. To these
belong strong electron correlation giving rise to Coulomb
blockade and Kondo physics, electron-phonon interactions responsible
for device heating and measurable in inelastic tunneling spectra, as
well as electron-photon interactions for molecular
photo-switches \cite{Cuevas2010}. While a qualitative understanding of these complex
processes has been obtained in many cases, quantitative agreement
between first-principles theory and experiment is still unsatisfactory
for the seemingly simple case of ballistic steady state transport \cite{Lindsay2007}.

The majority of such studies for realistic devices are based on the nonequilibrium Green's
function (NEGF) technique
\cite{Kadanoff1989}, which reduces to the traditional Landauer
transmission formalism \cite{Landauer1957,Buttiker1986} for coherent transport. The device Green's function is
then constructed from the Kohn-Sham (KS) single-particle Hamiltonian of ground state Density
Functional Theory (DFT) including the effects of the leads through suitable
self-energies. For finite bias, more sophisticated approaches determine
the actual potential profile on the molecule by solving the Poisson
equation in order to achieve a self-consistent treatment of
Hamiltonian and Green's function
\cite{Cuevas2010}.

In recent years, also time-dependent Density
Functional Theory (TDDFT) has become popular in the context of
molecular electronics
\cite{Tomfohr2001,Baer2004,Burke2005,Bushong2005,Kurth2005,Cheng2006,Zheng2007,Evans2009}. The reason for this is twofold. First, TDDFT
goes beyond the {\em ad-hoc} application of the DFT Hamiltonian for
non-equilibrium systems and provides a more rigorous theoretical
foundation \cite{Stefanucci2004}. In fact, calculations at this level provide corrections to
the conventional DFT-NEGF results for steady state currents, provided
non-local exchange-correlation functionals are employed \cite{Evers2004,Sai2005,Vignale2009,Yam2011}. Second, TDDFT
allows for the description of dynamical processes like the switching
behavior of molecular devices and alternating currents. The formalism
may also be easily extended to cover the interaction with light, which
paves the way to study photo-assisted and photo-suppressed transport,
the fluorescence of contacted molecules or atomistic opto-electronic
devices like photoswitches.

So far, several implementations of TDDFT for open systems have been suggested,
which may be classified according to the way the transport boundary conditions are treated. The
simplest approach is to approximate the mesoscopically large
leads by atomic clusters of finite size \cite{Tomfohr2001,Bushong2005,Cheng2006,Evans2009}. The Kohn-Sham orbitals are
then time-propagated under the influence of a potential difference between
left and right leads and changes of molecular charges in the contacts
allow one to quantify the current. This approach has the advantage
that TDDFT algorithms for isolated systems can be applied to the
transport scenario without large changes. On the downside, a steady-state current is only
transiently reached due to charge accumulation. Also, unphysical
oscillations of the current never  fully die out, because of the discrete level spectrum of the
contacts.

Another path was pursued by Burke and
co-workers \cite{Burke2005,Koentopp2008}. Here a ring-topology of the electronic
circuit is used to
avoid the conceptual difficulty of having two different chemical
potentials as in the Landauer picture. A constant electric field on
the ring is used to generate a current and coupling to a heat bath
within a master equation approach allows for the establishment of a
steady-state.

\begin{figure}
\begin{center}
\includegraphics[angle=0,scale=0.25]{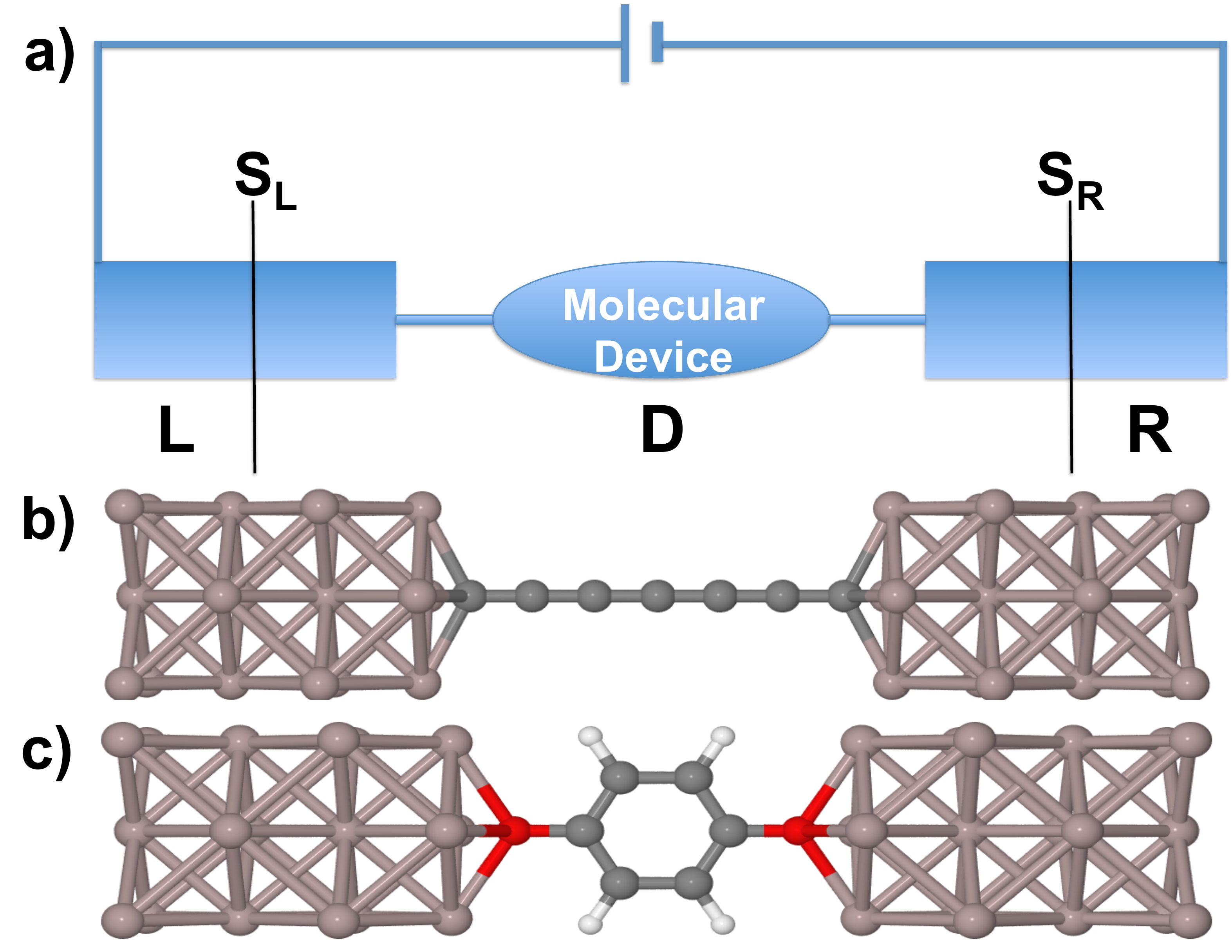}
\caption{(a) Schematic device setup with D denoting the central device
  region. The left (L) and right (R) leads are assumed to be semi-infinite and in
  thermal equilibrium. (b) Seven-membered carbon chain between Al nanowires in (001) direction.
Frontier carbon atoms in hollow position of the Al surface at 1.0 \AA\  distance.
Bond lengths (\AA): C-C = 1.32, Al-Al = 2.86.
(c) 1,4-benezenediol molecule inbetween Al nanowires.}
\label{LDR}
\end{center}
\end{figure}

A third approach is given by a separation of the conductor into a
finite device region and two semi-infinite leads, very similar to the
treatment within the standard DFT-NEGF approach (see Fig.~\ref{LDR}a). The time-dependent
Kohn-Sham equations need to be solved for the central region
only and the effect of the leads is exactly accounted for by properly
defined self-energies. Several model studies  along these lines
are documented in the literature
\cite{Kurth2005,Stefanucci2006,Stefanucci2008,Khosravi2009}.
 Finally, a closely related path is followed
 by Chen and co-workers \cite{Zheng2007,Yam2008,Zheng2010}. Instead of propagating Kohn-Sham states,
the equations of motion are written in terms of the reduced density
matrix for the device region.  This fact allows one to obtain the initial conditions from a
straightforward application of equilibrium Green's function theory.

For all of the methods mentioned above, simulations on realistic
molecular devices which take the atomistic structure of both molecule
and leads into account are numerically demanding. This is because the
electron motion has to be fully resolved, leading to time steps in
the sub-fs range. Although only the central part of the device is
out of equilibrium, this region has to include several layers of the
lead material in addition to the functional molecule in question,
which raises the dimension of the Hamiltonian.  Only
in this way a smooth transition of potentials at the device-lead
interface is guaranteed. Moreover, comprehensive studies require
numerous simulations involving the variation of different  parameters, like different molecular binding
configurations, temporal bias profiles or light frequency and
amplitude in case of photo-induced processes. Hence, a computationally
efficient, but still predictive method based on TDDFT seems to be desirable.

In this contribution we present efforts in this direction and discuss
an implementation of the density functional based tight-binding method
(DFTB) \cite{Seifert1986,porezag1995ctb,elstner1998scc}
for open systems. The ground state DFTB approach is characterized by
several additional approximations beyond a particular exchange-correlation functional and
has already been generalized to the time domain (TD-DFTB, \cite{Niehaus2001a,Niehaus2005,Niehaus2009}) in the
spirit of TDDFT. Here we follow the mentioned density matrix approach
of Chen and co-workers to arrive at a fast simulation method for open systems driven by time dependent
bias potentials. In Sec.~\ref{secrdm} the basic equations of motion are
reviewed, while Sec.~\ref{DFTB} discusses the approximations underlying
the DFTB scheme and the necessary adaptions in the present
context. Several case studies are presented in Sec.~\ref{results} followed
by a brief summary and outlook in Sec.~\ref{summ}.

\section{Equation of motion for the reduced density matrix}
\label{secrdm}
\subsection{General formulas}
We consider a system as depicted in Fig.~\ref{LDR}a and aim at
evaluating the current through the central
device region (D) which is driven by a time-dependent bias potential
$V(t)$. The left (L) and right (R) lead are thought to extend from the
device region to infinity and feature a continuous density of states to
ensure proper dissipation.  In second quantization, the electronic
Hamiltonian of the whole system can be written as
\begin{equation}
\mathcal{H_\text{tot}}=\sum_{mn}H_{mn}d_{m}^{\dag}d_{n}+\sum_{\alpha,k}\varepsilon_{\alpha,k}
c_{\alpha,k}^{\dag}c_{\alpha,k}+\sum_{\alpha,k;m}(T_{\alpha,k;m}c_{\alpha,k}^{\dag}d_{m}+
\text{h.c.}).\label{Hamiltonian}
\end{equation}
Here, the first term describes the
device region, where $d_{m}^{\dag}$($d_{m}$) is the creation
(annihilation) operator for an electron in the atomic orbital  $m$,
and $H_{mn}$ stands for the
corresponding Hamiltonian in this basis. The second term describes the
contacts ($\alpha \in $ \{L, R\}), where $c_{\alpha,k}^{\dag}$($c_{\alpha,k}$) is the
creation (annihilation) operator for an electron in the single-particle
state $k$ in the $\alpha$th contact,
while $\varepsilon_{\alpha,k}$ is the corresponding  level energy. The
third term describes the coupling
between device region and contacts and features the
hopping matrix $\boldsymbol{T}$ (matrices will be denoted by bold face letters). Although the Hamiltonian in
Eq.~(\ref{Hamiltonian}) is formally valid only for non-interacting
particles, the yet to be specified parameters may effectively account
for electron-electron interactions, for example in a DFT context. Moreover,
a self consistent determination of the device Hamiltonian and hopping
matrix will result in a time dependence of these quantities for
varying external bias.

Next, we define the lesser Green's
function \cite{Haug2008} of the device region as
\begin{equation}
G_{mn}^{<}(t,t')= i \langle d_{n}^{\dag}(t')d_{m}(t)\rangle,\label{Gless}
\end{equation}
and apply the equation of motion for Heisenberg operators to arrive at
\begin{equation}
i\frac{\partial}{\partial t}\boldsymbol{G}^{<}(t,t')= [\boldsymbol{H},\boldsymbol{G}^{<}(t,t')] + \sum_{\alpha}\boldsymbol{Q}_{\alpha}(t,t'),\label{KBE}
\end{equation}

where $\boldsymbol{Q}_{\alpha}(t,t')$ is given in terms of the
coupling matrix and the lead-device lesser Green's function
\begin{equation}
  G_{\alpha,k;n}^{<}(t,t')= i \langle d_{n}^{\dag}(t')c_{\alpha,k}(t)\rangle
\end{equation}
as follows
\begin{equation}
Q_{\alpha,mn}(t,t')= \sum_{k}[T_{m;\alpha,k}(t)G_{\alpha,k;n}^{<}(t,t')
-G_{m;\alpha,k}^{<}(t,t')T_{\alpha,k;n}(t')].\label{Qterm1}
\end{equation}

Following the standard NEGF technique \cite{Haug2008}, the
term $\boldsymbol{Q}_{\alpha}(t)$ can be further rewritten in a form that
incorporates only matrices with the dimension of the device region:
\begin{eqnarray}
\boldsymbol{Q}_{\alpha}(t,t')&=&+\int_{-\infty}^{+\infty}d\tau[\boldsymbol{\Sigma}_{\alpha}^{r}(t,\tau)\boldsymbol{G}^{<}(\tau,t')+
\boldsymbol{\Sigma}_{\alpha}^{<}(t,\tau)\boldsymbol{G}^{a}(\tau,t')]\nonumber\\
&&-\int_{-\infty}^{+\infty}d\tau[\boldsymbol{G}^{<}(t,\tau)\boldsymbol{\Sigma}_{\alpha}^{a}(\tau,t')
+\boldsymbol{G}^{r}(t,\tau)\boldsymbol{\Sigma}_{\alpha}^{<}(\tau,t')].\label{Qterm2}
\end{eqnarray}
Here, $\boldsymbol{G}^{r(a)}$ are the retarded (advanced) Green's functions of the
device region, which satisfy the Dyson equation
\begin{eqnarray}
&&(i\frac{\partial}{\partial t}-\boldsymbol{H}(t))\boldsymbol{G}^{r(a)}(t,t')\nonumber\\
&-&\sum_{\alpha}\int_{-\infty}^{+\infty}d\tau\boldsymbol{\Sigma}_{\alpha}^{r(a)}(t,\tau)\boldsymbol{G}^{r(a)}(\tau,t')
=\delta(t-t'),\label{dyson}
\end{eqnarray}
while $\boldsymbol{\Sigma}_{\alpha}^{r,a,<}$ are the
retarded (advanced, lesser) self-energies
due to the coupling to the  $\alpha$th contact. They are given as
\begin{eqnarray}
\boldsymbol{\Sigma}_{\alpha}^{r,a,<}(t,\tau)=\boldsymbol{T}_{\alpha}^{\dag}(t)\boldsymbol{g}_{\alpha}^{r,a,<}(t,\tau)\boldsymbol{T}_{\alpha}(\tau),\label{Self-E}
\end{eqnarray}
with $\boldsymbol{g}_{\alpha}^{r,a,<}$ denoting the Green's functions of the $\alpha$th
contact.

Equations (\ref{KBE},\ref{Qterm2},\ref{dyson},\ref{Self-E}) form
a closed set of equations to describe the dynamical properties of
non-equilibrium systems and will be the theoretical basis of our
method. Instead of working with the full lesser Green's function, a significant simplification is achieved by
turning to the reduced density matrix for the device region ($\boldsymbol{\sigma}(t)$)
as the key quantity, which is dependent on a single time argument only.
The quantity $\boldsymbol{\sigma}(t)$ is directly
related to the lesser Green's function $\boldsymbol{G}^{<}(t,t')$
\begin{equation}
\boldsymbol{\sigma}(t)=-i\boldsymbol{G}^{<}(t,t).\label{Definition}
\end{equation}
Thus, the equation of motion for $\boldsymbol{\sigma}(t)$ can be obtained from equation
(\ref{KBE}) as
\begin{equation}
i\frac{\partial}{\partial
t}\boldsymbol{\sigma}(t)=[\boldsymbol{H}(t),\boldsymbol{\sigma}(t)]-i\sum_{\alpha}\boldsymbol{Q}_{\alpha}(t).\label{RDM}
\end{equation}
For vanishing coupling (i.e. $\boldsymbol{T}=0$), $\boldsymbol{Q}$ vanishes and
equation (\ref{RDM}) reduces to the conventional quantum Liouville equation for
closed systems. The quantity $\boldsymbol{Q}_{\alpha}$ describes dissipation
effects due to the presence of the semi-infinite contacts and gives
rise to finite lifetimes of states located in the central region. It is also noted that the trace
of the matrix $\boldsymbol{Q}_{\alpha}$ gives the particle current
through the interface $S_\alpha$ (see Fig.~\ref{LDR}a) in contact $\alpha$ \cite{Zheng2007}
\begin{equation}
I_{\alpha}(t)=-\text{Tr}[\boldsymbol{Q}_{\alpha}(t)].\label{current}
\end{equation}

\subsection{Wide-band approximation for the dissipation term $\boldsymbol{Q}_\alpha$}
Direct application of Eq.~(\ref{Qterm2}) to evaluate the dissipation
term $\boldsymbol{Q}_\alpha$ turns out to be computationally
demanding. This is due to the necessity to store and integrate over
two-time quantities like the self-energies. A common solution is
to invoke the wide band approximation (WBA), where both the lead density of
states and the device-lead coupling are assumed to be smooth and not
strongly dependent on energy. This is a reasonable approximation for
simple metals in the linear response regime but fails for larger
applied bias and/or leads with a more complex density of states, like
for example nanotubes. The WBA is however not without alternative in
the present context. Recently, Zheng and co-workers presented a
hierarchical equation of motion approach, which goes beyond the WBA at
a tolerable increase in computational effort \cite{Zheng2010}.

Within the WBA, the self-energies
$\boldsymbol{\Sigma}^{r/a}_\alpha$ become local in time and can be
written as \cite{Zheng2007}
\begin{eqnarray}
\boldsymbol{\Sigma}_{\alpha}^{r,a}(t,t')&=&\left(\boldsymbol{\Gamma_\alpha} \mp  i \boldsymbol{\Lambda_\alpha}\right)\delta(t-t').\label{WBLSelfR}
\end{eqnarray}

Here, the matrices $\boldsymbol{\Gamma_\alpha}$ and $\boldsymbol{\Lambda_\alpha}$ denote
the hermitian and anti-hermitian part of
$\boldsymbol{\Sigma}^{a}_\alpha(E)$, respectively\footnote{Note
  that the definition of $\boldsymbol{\Gamma_\alpha}$ and
  $\boldsymbol{\Lambda_\alpha}$ is often interchanged in the literature. We
  keep with the notation of reference \cite{Zheng2007} to ease comparison.}, and are evaluated
at the common Fermi energy ($E_F$) of the unbiased system. Assuming
that the non-interacting
electrons in contact $\alpha$ are in local equilibrium characterized  by
a Fermi-Dirac distribution function $f^\alpha(\epsilon)$ with
chemical potential $\mu_0$ and assuming further
that the bias potential $V_\alpha(t)$ leads to a rigid shift of the
lead energy levels, one obtains
\begin{eqnarray}
\boldsymbol{\Sigma}_{\alpha}^{<}(t,t')&=&\frac{2i}{\pi}\boldsymbol{\Lambda}_{\alpha}
\exp\left\{i\int_{t'}^{t}d\tau V_{\alpha}(\tau)\right\} \int_{-\infty}^{+\infty}d\epsilon
f^{\alpha}(\epsilon)e^{-i\epsilon(t-t')}.\label{WBLSelfL}
\end{eqnarray}

With the expressions (\ref{WBLSelfR},\ref{WBLSelfL}) above, the
dissipation term $\boldsymbol{Q}_{\alpha}(t)$ can be simplified to \cite{Zheng2007}
\begin{equation}
\boldsymbol{Q}_{\alpha}(t)=i[\boldsymbol{\Gamma}_{\alpha},\boldsymbol{\sigma}(t)]+\{\boldsymbol{\Lambda}_{\alpha},\boldsymbol{\sigma}(t)\}+\boldsymbol{K}_{\alpha}(t).\label{WBLQ}
\end{equation}
Here, the first term represents the renormalization of the device energy levels due to the
$\alpha$th contact and involves a commutator; the second term
involving an anti-commutator describes the level
broadening, while the hermitian matrix  $\boldsymbol{K}_{\alpha}(t)$
\begin{eqnarray}
\boldsymbol{K}_{\alpha}(t)&=&-\frac{2i}{\pi}\boldsymbol{U}_{\alpha}(t)\int_{-\infty}^{\mu_{0}}\frac{d\epsilon\,
e^{i\epsilon t}}{\epsilon\boldsymbol{I} -\boldsymbol{H}(0)-\boldsymbol{\Sigma}^{r}}\boldsymbol{\Lambda}_{\alpha}\nonumber\\
&&-\frac{2i}{\pi}\int_{-\infty}^{\mu_{0}}[\boldsymbol{I}-\boldsymbol{U}_{\alpha}(t)e^{i\epsilon
t}]\frac{d\epsilon}{\left(\epsilon-V_{\alpha}(t)\right) \boldsymbol{I} -\boldsymbol{H}(t)-\boldsymbol{\Sigma}^{r}}\boldsymbol{\Lambda}_{\alpha}+h.c.\label{Kterm}
\end{eqnarray}

incorporates the history of the applied bias via the propagator
 $\boldsymbol{U}_{\alpha}(t)$
\begin{equation}
\boldsymbol{U}_{\alpha}(t)=\exp\left\{-i\int_{0}^{t}[\boldsymbol{H}(\tau)+\boldsymbol{\Sigma}^{r}-\boldsymbol{I}V_{\alpha}(\tau)]d\tau\right\}\label{Uterm}
\end{equation}
The term $\boldsymbol{\Sigma}^{r}$ is a summation of the retarded self-energy functions
for both contacts.

Equations (\ref{RDM}, \ref{WBLQ}, \ref{Kterm}, \ref{Uterm}) provide an
approximate approach to calculate the transient non-equilibrium
density matrix for open systems. Zheng {\em et.~al.} have combined
the scheme described above with TDDFT to simulate the
time-dependent transport in nanoscale systems \cite{Zheng2007,Yam2008}. Next, we
will describe how the density matrix $\boldsymbol{\sigma}(t)$, Hamiltonian $\boldsymbol{H}(t)$, and
dissipation term $\boldsymbol{Q}_{\alpha}(t)$ are constructed in the framework of
the density-functional based tight-binding (DFTB) method, which aims to
provide a more efficient but still accurate method for real
material simulations.

\section{Density Functional based Tight-Binding Method}
\label{DFTB}
\subsection{DFTB for closed systems}
As an approximate DFT method, DFTB was initially developed
to study the ground state of finite or periodic systems. Details on
the derivation and performance for these class of materials may be found
in several reviews \cite{frauenheim2002asc,Seifert2007,otte2007lsc}. As in conventional
tight-binding models, the total energy of DFTB consists of a band
structure part $E_\text{BS}$ and a repulsive energy part $E_\text{rep}$,
\begin{equation}
E^\text{DFTB}=E_\text{BS}+E_\text{rep}.\label{TotalE}
\end{equation}
Since we will not consider structural changes
during electron transport in the present work, only the band structure
term is relevant for the further discussion. It takes the form

\begin{equation}
E_\text{BS}=\sum_{i}^\text{occ}\langle\Psi_{i}|H_{0}|\Psi_{i}\rangle
+\frac{1}{2} \sum_{\alpha,\beta}\gamma_{\alpha\beta}\Delta
q_{\alpha}\Delta q_{\beta},\label{EBS}
\end{equation}

which is obtained through an expansion of the DFT total energy around
a reference density $n^{0}(\boldsymbol{r})$ up to second order \cite{elstner1998scc}.  The latter is taken to be
a superposition of the neutral densities of all atoms comprising the
system in question ($n^{0}(\boldsymbol{r})=\sum_\alpha n^{0}_\alpha(\boldsymbol{r})$).
The first term in Eq.~(\ref{EBS}) involves a sum over expectation values
of the DFT Hamiltonian evaluated at $n^{0}$:
\begin{equation}
H_{0}(\boldsymbol{r})=-\frac{\nabla^{2}}{2}+V_{ext}(\boldsymbol{r})+\int
d\boldsymbol{r}'\frac{n_{0}(\boldsymbol{r}')}{|\boldsymbol{r}-\boldsymbol{r}'|}+V_\text{XC}[n_{0}],\label{DFT}
\end{equation}
with the kinetic energy operator,  the  external potential, the Hartree potential
generated by $n_{0}$ and the  DFT exchange-correlation potential.

The last quantity  in  Eq.~(\ref{EBS}) represents the second-order
term in the mentioned density expansion and accounts for charge
transfer effects. Here, $\Delta q_{\alpha}$ is
defined as the difference of the Mulliken charges between charged
and neutral atom $\alpha$, {\em i.e.} $\Delta q_{\alpha}=q_{\alpha}-q_{\alpha}^{0}$. The matrix
$\gamma_{\alpha\beta}$ is a measure of the electron-electron interaction and
decays like $1/|\boldsymbol{R}_{\alpha}-\boldsymbol{R}_{\beta}|$ for large distances
between atoms at $\boldsymbol{R}_{\alpha}$ and $\boldsymbol{R}_{\beta}$. For
the on-site case, the Hubbard-like
parameter $U_\alpha=\gamma_{\alpha\alpha}$ is taken from full atomic DFT
calculations and represents the chemical hardness of the respective
element. An interpolation formula between these limiting cases  can be
analytically constructed  \cite{elstner1998scc} by invoking a monopole approximation
for the density fluctuations $\delta n_\alpha$, which are assumed to take the form
\begin{equation}
\delta n_{\alpha}(\boldsymbol{r})=  \frac{ \Delta
  q_{\alpha}\tau_{\alpha}^{3}}{8\pi}e^{-\tau_{\alpha}|\boldsymbol{r}-\boldsymbol{R}_{\alpha}|}\text{,
  with\,\,} \int d{\bf r}\,\, \delta n_{\alpha}(\boldsymbol{r}) = \Delta
  q_{\alpha}.  \label{MomoPole}
\end{equation}
Here, the effective decay constant $\tau_\alpha=\frac{16}{5} U_{\alpha}$ is determined by the constraint $U_\alpha=\gamma_{\alpha\alpha}$.

In the practical implementation of DFTB, the Kohn-Sham orbitals
$|\Psi_{i}\rangle$ are expanded in a set of localized atomic orbitals (AO)
$\{|\phi_\mu\rangle\}$ that are again obtained from a full atomic DFT
calculation
\begin{equation}
|\Psi_{i}\rangle=\sum_{\mu}c_{i\mu}|\phi_\mu\rangle.
\end{equation}
Usually only occupied AO are used in the molecular
expansion, but this is not a necessary requirement of the model. The atomic DFT calculations are converged
with respect to the basis set size, so that each AO is constructed from a superposition
of many Slater-type orbitals. Furthermore, an additional confinement
potential  in the atomic calculations leads to a suppression of the
long-range tail of the AO, making them suitable expansion functions
for the molecular case \cite{porezag1995ctb}. The minimal basis set of DFTB is expected to
be more accurate than conventional contracted basis sets of the same
size, e.g.~Pople's STO-3G.

In a two-center approximation, matrix elements of the DFT Hamiltonian
in Eq.~(\ref{DFT}) ($H^0_{\mu\nu}$) are numerically computed and stored in
Slater-Koster \cite{Slater1954} tables as function of the inter-element
distance. The Mulliken charges used to estimate the second-order term
in the density expansion are given in terms of the molecular orbital
coefficients ($c_{i\mu}$) and the overlap of two atomic orbitals
($S_{\mu\nu}= \langle \phi_\mu | \phi_\nu \rangle$):

\begin{eqnarray}
q_{\alpha}&=&\text{Tr}_{\alpha}[\boldsymbol{\sigma S}]\nonumber\\
&=&\frac{1}{2}\sum_{i}^{occ}\sum_{\mu\in
\alpha}\sum_{\nu}(c_{i\mu}^{*}c_{i\nu}S_{\mu\nu}+c_{i\nu}^{*}c_{i\mu}S_{\nu\mu}).\label{MulliQ}
\end{eqnarray}
The trace is performed over all atomic orbitals on site $\alpha$
and $\boldsymbol{\sigma}$ denotes the density matrix in the AO basis. Thus the total energy of DFTB
will be a function of the expansion
coefficients $\{c_{i\mu},c_{j\nu}^{*}\}$ in the given basis set. The
ground state energy  can be obtained from the variational principle
\begin{equation}
\delta
(E_{BS}-\sum_{i}\epsilon_{i}c_{i\mu}^{*}c_{i\nu}S_{\mu\nu})/\delta
c_{j\mu}^{*}=0,\label{VariE}
\end{equation}
where the $\epsilon_{i}$ are Lagrange multipliers.

The solution of equation (\ref{VariE}) gives rise to the secular
equation
\begin{equation}
\sum_{\nu}(H_{\mu\nu}-\epsilon_{i}S_{\mu\nu})c_{i\nu}=0,\label{SecEqu}
\end{equation}
where the Hamiltonian matrix element $H_{\mu\nu}$ for atom pair
$(\alpha,\beta)$ reads
\begin{equation}
H_{\mu\nu}=H_{\mu\nu}^{0}+\frac{1}{2}S_{\mu\nu}\sum_{\gamma}(\gamma_{\alpha\gamma}+\gamma_{\beta\gamma})\Delta
q_{\gamma}\quad \mu\in \alpha, \nu \in \beta.\label{Hamil}
\end{equation}

Equation (\ref{SecEqu}) can be viewed as an approximate version of
the Kohn-Sham equation in DFT. The second term in (\ref{Hamil})
accounts for changes of the electron density with respect to the
reference ($n^{0}$) and needs to be determined self-consistently via
repeated evaluation of Eqs.~(\ref{SecEqu}, \ref{MulliQ}, and
\ref{Hamil}), very similar to self consistent field approaches. The
computational efficiency of the DFTB method grounds on two
facts. First, the minimal but still accurate basis set is smaller
then in conventional DFT approaches and second, the construction of
the Hamiltonian requires little CPU time due the precomputation of all
numerical integrals.

\subsection{DFTB for open systems}
For open systems, the concept of a total energy is not
well-defined, and the variational principle and secular equation
do not hold. Thus, the DFTB method needs to be generalized in oder to deal
with a semi-infinite system out of equilibrium. Recalling the equations
(\ref{RDM},\ref{WBLQ},\ref{Kterm},\ref{Uterm}) in Sec.~\ref{secrdm} which
describe the transient non-equilibrium dynamics in open systems, the
key ingredients which determine the reduced density matrix $\boldsymbol{\sigma}(t)$
are the Hamiltonian $\boldsymbol{H}(t)$ and self-energy functions
$\boldsymbol{\Sigma}_{\alpha}$. Here we present how to construct these matrices in
the framework of DFTB.

\subsubsection{Device Hamiltonian}
In the spirit of DFTB, the Hamiltonian should take the general form
\begin{equation}
H_{\mu\nu}=H^{0}_{\mu\nu}[n_{0}]+\delta V_{\mu\nu}[\{\Delta q_{\alpha}\}]\label{HDFTB}
\end{equation}
where the first term is the zero-order DFTB Hamiltonian,  while the second term includes the Hartree potential $\delta
V_{H}$ and exchange-correlation potential $\delta V_{XC}$ due to
the difference density $\delta n = \sum_\alpha \delta n_\alpha$.

For closed systems, where the electrostatic potential can be assumed
to tend to zero at infinity, the Hartree potential $\delta V_{H}(\boldsymbol{r})$
is given by
\begin{equation}
\delta V_{H}(\boldsymbol{r})=\int d\boldsymbol{r}'\frac{\delta
n(\boldsymbol{r}')}{|\boldsymbol{r}-\boldsymbol{r}'|}.\label{HartreeP}
\end{equation}
Using Eq.~(\ref{MomoPole}), this results in the $\gamma$-matrix dependent
second-oder term of the Hamiltonian in Eq.~(\ref{Hamil}). For an
electronic device like the one depicted in Fig.~\ref{LDR}a, the
electrostatic potential in the left and right contact is set by the
applied bias. The potential in the central region must therefore be
obtained from the Poisson equation
\begin{equation}
\nabla^{2}\delta V_{H}(\boldsymbol{r})=-4\pi\delta n(\boldsymbol{r})
\label{Poisson}
\end{equation}
under these boundary conditions. In our approach, the device density
matrix $\boldsymbol{\sigma}(t)$ is used to compute time-dependent Mulliken
charges according to Eq.~(\ref{MulliQ}), which are used to set up the
difference density $\delta n(\boldsymbol{r})$ via
Eq.~(\ref{MomoPole}). Taking $\tau_{\alpha}=\frac{16}{5}U_{\alpha}$ as
above, some exchange and correlation effects are taken into account at
this point, since the Hubbard-like parameters $U_{\alpha}$ stem from a
DFT calculation and are considerably smaller than unscreened
Hartree-only parameters \cite{Niehaus2005a}.

The Poisson
equation is then discretized on a grid and solved in real space for a
sufficiently large box enclosing the device region. Note
that an inherent assumption of such an approach is that the potential
at the device-lead interface equals the potential in the contact
interior. In other words, the device region should include enough layers of
the contact material to ensure efficient screening.

Similar to the electron density, $\delta V_H(\boldsymbol{r})$ is projected
on the atomic sites through
\begin{equation}
\delta V_{\alpha}=\frac{1}{\lambda}\int_\text{box} d\boldsymbol{r} \,\delta
V_H(\boldsymbol{r}) e^{-\tau_{\alpha}|\boldsymbol{r}-\boldsymbol{R}_{\alpha}|}\label{DeltaVr}
\end{equation}
with normalization factor $\lambda=\int_\text{box}
d\boldsymbol{r}e^{-\tau_{\alpha}|\boldsymbol{r}-\boldsymbol{R}_{\alpha}|}$. The
required second term in the device Hamiltonian Eq.~(\ref{HDFTB}) may
now be evaluated as
\begin{equation}
\delta V_{\mu\nu}=\frac{1}{2}(\delta V_{\alpha}+\delta
V_{\beta})S_{\mu\nu},\quad \mu\in \alpha, \nu \in \beta.\label{DeltaV}
\end{equation}
The approach discussed in this subsection follows the work of Pecchia
{\em et.~al.} \cite{Pecchia2008}, who earlier implemented the static Landauer formalism for the
DFTB method.

\subsubsection{Self-energy}
From Eq.~(\ref{Self-E}), the retarded self-energy function
$\boldsymbol{\Sigma}_{\alpha}^{r}$  can be expressed in the energy domain as
\begin{equation}
\boldsymbol{\Sigma}_{\alpha}^{r}(E)=\boldsymbol{T}_{\alpha}^{\dag}\boldsymbol{g}_{\alpha}^{r}(E)\boldsymbol{T}_{\alpha},\label{Self-E2}
\end{equation}
where $\boldsymbol{g}_{\alpha}^{r}$ is the retarded Green's function of the
$\alpha$th contact. As in previous work, we assume that the contacts are
semi-infinite periodic lattices, which are divided into principle
layers (PLs) along the transport direction. The PLs are
chosen so wide that only interactions between nearest PLs need to
be considered. Then the coupling matrix $\boldsymbol{T}_{\alpha}$ between
contact and device region (which includes at least one PL in
the device part) will be restricted to one PL, and only the surface
block of $\boldsymbol{g}_{\alpha}^{r}$, the {\em surface} Green's function $\boldsymbol{g}_{\alpha}^{s}(E)$,
is needed to calculate $\boldsymbol{\Sigma}_{\alpha}^{r}$.

The surface Green's function $\boldsymbol{g}_{\alpha}^{s}(E)$ can be obtained from
the standard renormalization method \cite{LopezSancho1985}. Denoting $\boldsymbol{H}_{00}$
as the Hamiltonian of one PL, and $\boldsymbol{H}_{01}$ as the coupling matrix
between the nearest PL, $\boldsymbol{g}_{\alpha}^{s}$ is given by
\begin{equation}
\boldsymbol{g}_{\alpha}^{s}(E)=(E\boldsymbol{I}-\boldsymbol{\zeta}_{\infty})^{-1}\label{SGF}
\end{equation}
where the matrices $\boldsymbol{\zeta}_{i}$ are calculated from the recurrence
relations
\begin{eqnarray}
\boldsymbol{\zeta}_{i+1}&=&\boldsymbol{\zeta}_{i}+\boldsymbol{a}_{i}(E\boldsymbol{I}-\boldsymbol{\eta}_{i})^{-1}\boldsymbol{b}_{i}\label{zeta}\\
\boldsymbol{\eta}_{i+1}&=&\boldsymbol{\eta}_{i}+\boldsymbol{a}_{i}(E\boldsymbol{I}-\boldsymbol{\eta}_{i})^{-1}\boldsymbol{b}_{i}+\boldsymbol{b}_{i}(E\boldsymbol{I}-\boldsymbol{\eta}_{i})^{-1}\boldsymbol{a}_{i}\label{zeta2}\\
\boldsymbol{a}_{i+1}&=&\boldsymbol{a}_{i}(E\boldsymbol{I}-\boldsymbol{\eta}_{i})^{-1}\boldsymbol{a}_{i}\label{a}\\
\boldsymbol{b}_{i+1}&=&\boldsymbol{b}_{i}(E\boldsymbol{I}-\boldsymbol{\eta}_{i})^{-1}\boldsymbol{b}_{i}\label{b}
\end{eqnarray}
with the initial matrices
$\boldsymbol{\zeta}_{0}=\boldsymbol{H}_{\alpha,00},\boldsymbol{\eta}_{\alpha,0}=\boldsymbol{H}_{\alpha,00},\boldsymbol{a}_{0}=\boldsymbol{H}_{\alpha,01},\boldsymbol{b}_{0}=\boldsymbol{H}_{\alpha,01}^{\dag}$. In
the simplest approximation, the PL matrices
$\boldsymbol{H}_{\alpha,00}$ and $\boldsymbol{H}_{\alpha,01}$ are
obtained from a conventional DFTB calculation in which the contact is
treated as a finite cluster. Although the semi-infinite nature of the contacts
is still accounted for by the renormalization method, computational
artefacts have recently been reported for this approach \cite{Beste2008}. Instead, we
obtain the PL matrices from simulations using periodic boundary
conditions with converged Brillouin zone sampling \cite{Aradi2007a}.

After construction of the surface Green's function $\boldsymbol{g}_{\alpha}^{s}(E)$, it is
straightforward to compute the self-energy at the Fermi energy
\begin{equation}
\boldsymbol{\Sigma}_{\alpha}^{r}(E_F)=\boldsymbol{H}_{\alpha,01}^{\dag}\boldsymbol{g}_{\alpha}^{s}(E_F)\boldsymbol{H}_{\alpha,01},\label{Self-E3}
\end{equation}
which provides the necessary parameters for the  wide-band
approximation used to determine the dissipation term
$\boldsymbol{Q}_\alpha$.

\subsubsection{Initial density matrix }
We assume that the device is in thermal equilibrium at $t=0$ with a
common Fermi energy for the full lead-device-lead system. Only for
$t>0$ a bias is applied which drives the central region
out of equilibrium and results in a current flow. The initial
conditions for the solution of our central equation (\ref{RDM}) may
therefore be obtained from equilibrium Green's function theory. For
the reduced density matrix $\boldsymbol{\sigma}(0)$ we have

\begin{equation}
\boldsymbol{\sigma}(0)=\int_{-\infty}^{+\infty}\frac{dE}{2\pi}f(E)\Im\{
\boldsymbol{G}^{r}(E)\},\label{IMGR}
\end{equation}
where $\Im \{\boldsymbol{G}^{r}\}$ is the imaginary part of retarded Green's function
\begin{equation}
\boldsymbol{G}^{r}(E)=[E\boldsymbol{I}-\boldsymbol{H}(0)-\boldsymbol{\Sigma}^{r}(E)]^{-1}.\label{GRE}
\end{equation}
  Notice that $\boldsymbol{H}(0)$ is determined by $\boldsymbol{\sigma}(0)$ through equation
(\ref{HDFTB}) and (\ref{DeltaV}), thus the above equations need to be solved
self-consistently. Compared with the conventional DFTB scheme for closed systems,
equation ($\ref{IMGR}$) replaces the secular equation (\ref{SecEqu})
to obtain the density matrix, while equation ($\ref{HDFTB}$)
replaces equation (\ref{Hamil}) to construct the Hamiltonian.

After obtaining the initial conditions $\boldsymbol{\sigma}(0)$ and $\boldsymbol{H}(0)$, we use
the Runge-Kutta method to propagate the
 density matrix $\boldsymbol{\sigma}(t)$ with a time
step $\Delta t$ of 0.02 fs. Solution of the Poisson equation allows the Hamiltonian $\boldsymbol{H}(t)$
to be  updated according to equation ($\ref{HDFTB}$). In this approach,  $\boldsymbol{H}(t)$ depends only on the density at time $t$ and not on densities at former times, which corresponds to the adiabatic approximation for exchange and correlation. Finally, the dissipation
term $\boldsymbol{Q}_{\alpha}(t)$ is calculated from equation
($\ref{WBLQ}$) to compute the current and $\boldsymbol{\sigma}(t+\Delta t)$.  This completes the discussion on the implementation of
the TD-DFTB method for open systems.
\section{Test calculations}
\label{results}
The method described above has been applied to two prototypical
molecular junctions: a carbon chain with metallic transport
properties and an oxygen anchored  benzene molecule with small linear
response conductance. In Sec.~\ref{steadydev} we will discuss the
establishment of a steady state for asymptotically constant bias
potential, while Sec.~\ref{comp} is devoted to a comparison with {\em first
principles} TDDFT results in order to classify the accuracy of our
approximate scheme. 

\subsection{Development of steady state}
\label{steadydev}
The first system we study here is a carbon chain which bridges
two aluminum wires as shown
in Fig.~\ref{LDR}b. The chain consists of seven
carbon atoms with a distance of 1.32  ${\AA}$ between nearest
neighbors. The electrodes are given by Al nanowires of finite cross
sections oriented in (001) direction. The two end carbon atoms of the
chain are located at the
hollow sites of the surfaces, and the distance between the end
atom and the first layer of the Al surface is 1.0 ${\AA}$. Each PL in
the electrodes comprises 18 Al atoms. Besides the carbon chain,
two adjacent PLs are included in the device region amounting to 43
atoms in total. Similar
systems have been studied by several authors with the DFT-NEGF
method \cite{Larade2001, Brandbyge2002, Ke2004}. Thus, it serves as  a good test model for our
purposes here.

In an initial application of the TD-DFTB scheme, we apply a time
dependent bias potential that is exponentially turned on with a time
constant $T$ and  approaches a constant value $V_0$ asymptotically:
\begin{equation}
  \label{btd}
  V_{L}=0, \quad V_{R}(t)=V^0_R \left(1-e^{-t/T}\right).
\end{equation}

\begin{figure}
\begin{center}
\includegraphics[scale=0.75]{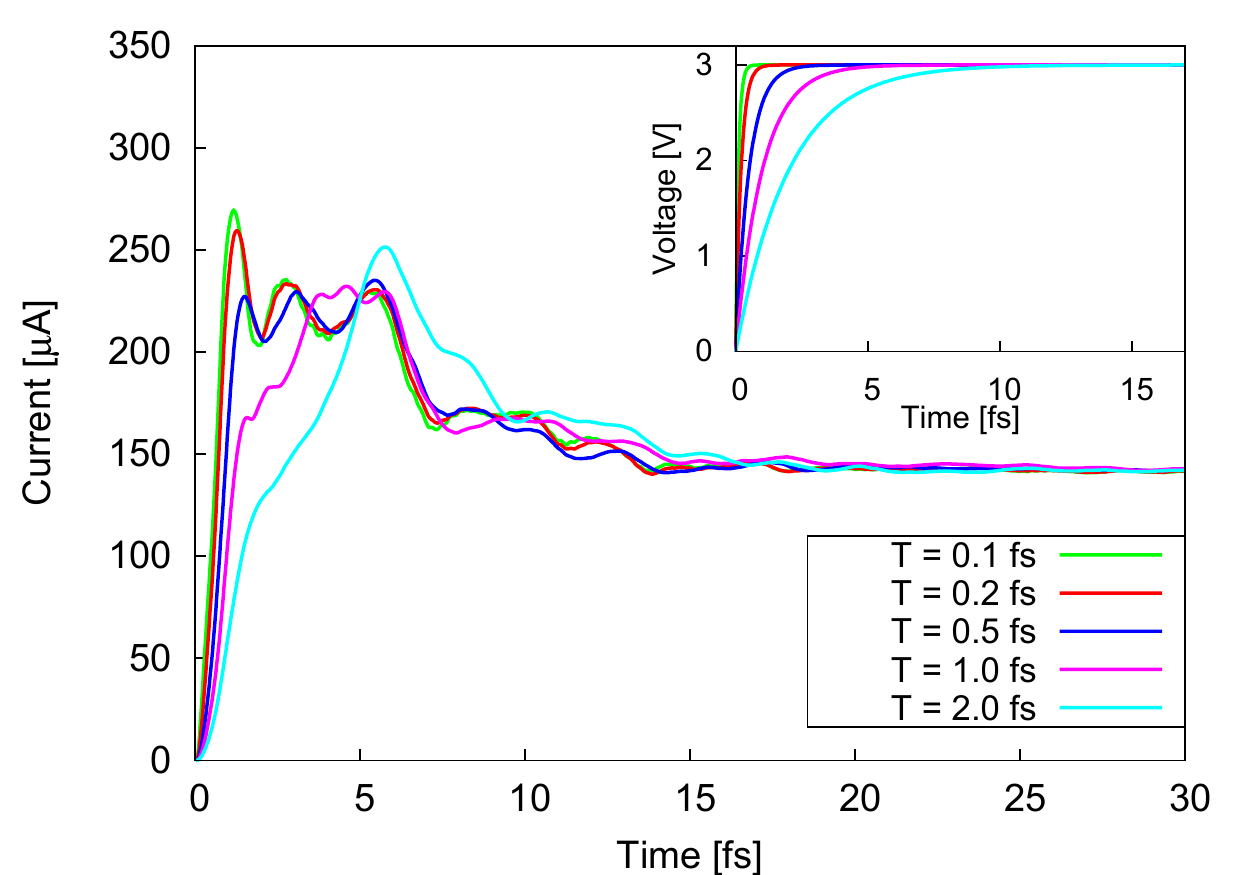}
\caption{TD-DFTB transient currents through the carbon chain of
  Fig.~\ref{LDR}b for an
  exponential turn on of the bias with different time constants
  $T$. The inset shows the time dependent bias approaching $V^0_R=$ 3 V}
\label{steady}
\end{center}
\end{figure}
Fig.~\ref{steady} depicts the current traces for various
time constants $T$. Although the initial transient currents differ
considerably, the same final steady state current is reached in all
cases. This is a nontrivial result, as earlier model studies indicated
the possibility of multiple steady state solutions \cite{Sanchez2006}, or even the
absence of a steady state \cite{Khosravi2008,Khosravi2009}. The latter investigations by Khosravi and
co-workers predict persistent current oscillations if the device
region supports a bound state which is uncoupled to the leads.

The existence of such undamped solutions may also be deduced in the
present formalism. If we assume that $\boldsymbol{\sigma}(t)$ and
$\boldsymbol{H}(t)$ approach constant values at $t_s$ such that
\begin{equation}
  \label{ts}
 \boldsymbol{\sigma}(t_s) \approx \boldsymbol{\sigma}(\infty) \wedge
 \boldsymbol{H}(t_s) \approx \boldsymbol{H}(\infty),
\end{equation}

the last part in the dissipation term $\boldsymbol{Q}_{\alpha}$ of Eq.~\ref{WBLQ} may be
decomposed into

\begin{equation}
\boldsymbol{K}_{\alpha}(t) = \boldsymbol{K}^\infty_{\alpha} +
\boldsymbol{\tilde{K}}_{\alpha}(t) \label{Kterm2}
\end{equation}

for $t>t_s$. Here $\boldsymbol{K}^\infty_{\alpha}$ is given by

\begin{eqnarray}
\boldsymbol{K}^\infty_{\alpha}&=&-\frac{2i}{\pi}\int_{-\infty}^{\mu_{0}}\frac{d\epsilon}{\left(\epsilon-V_\alpha^0\right)\boldsymbol{I}
  -\boldsymbol{H}(\infty)-\boldsymbol{\Sigma}^{r}}\boldsymbol{\Lambda}_{\alpha}+h.c. \label{Ktermsteady}
\end{eqnarray}

and contributes to the steady state current, while the component
\begin{eqnarray}
\boldsymbol{\tilde{K}}_{\alpha}(t)
&=&-\frac{2i}{\pi}\int_{-\infty}^{\mu_{0}}d\epsilon
 \exp\left[i\left\{ (\epsilon - V_\alpha^0)\boldsymbol{I} -
   \boldsymbol{H}(\infty)-\boldsymbol{\Sigma}^{r} \right\}
 (t-t_s)\right] \boldsymbol{\kappa}^\infty_\alpha(\epsilon)+h.c. \nonumber  \\
\text{with} && \nonumber
\\
\boldsymbol{\kappa}^\infty_\alpha(\epsilon) &=& e^{i\epsilon t_s} \boldsymbol{U}_{\alpha}(t_s)
 \left[ \frac{1}{\epsilon\boldsymbol{I}
  -\boldsymbol{H}(0)-\boldsymbol{\Sigma}^{r}} - \frac{1}{\left(\epsilon-V_\alpha^0\right)\boldsymbol{I}
  -\boldsymbol{H}(\infty)-\boldsymbol{\Sigma}^{r}}  \right]\boldsymbol{\Lambda}_{\alpha}\label{Ktermdyn}
\end{eqnarray}
might give rise to permanent oscillations of the current beyond
$t_s$. This occurs if at least one  of the eigenvalues of the effective
device Hamiltonian $\boldsymbol{H}+\boldsymbol{\Sigma}^{r}$ is purely
real, i.e.~a bound state exists which is completely uncoupled from the
contacts. As an additional trivial prerequisite, al least one
transmitting channel
must be coupled to the leads, otherwise $\boldsymbol{\Lambda}$ and
therefore also the current vanish.  In all of our practical simulations so far, we never encountered such
a situation of permanent current oscillations.
\subsection{Comparison with TDDFT}
\label{comp}
\begin{figure}
\begin{center}
\includegraphics[angle=0,scale=0.7]{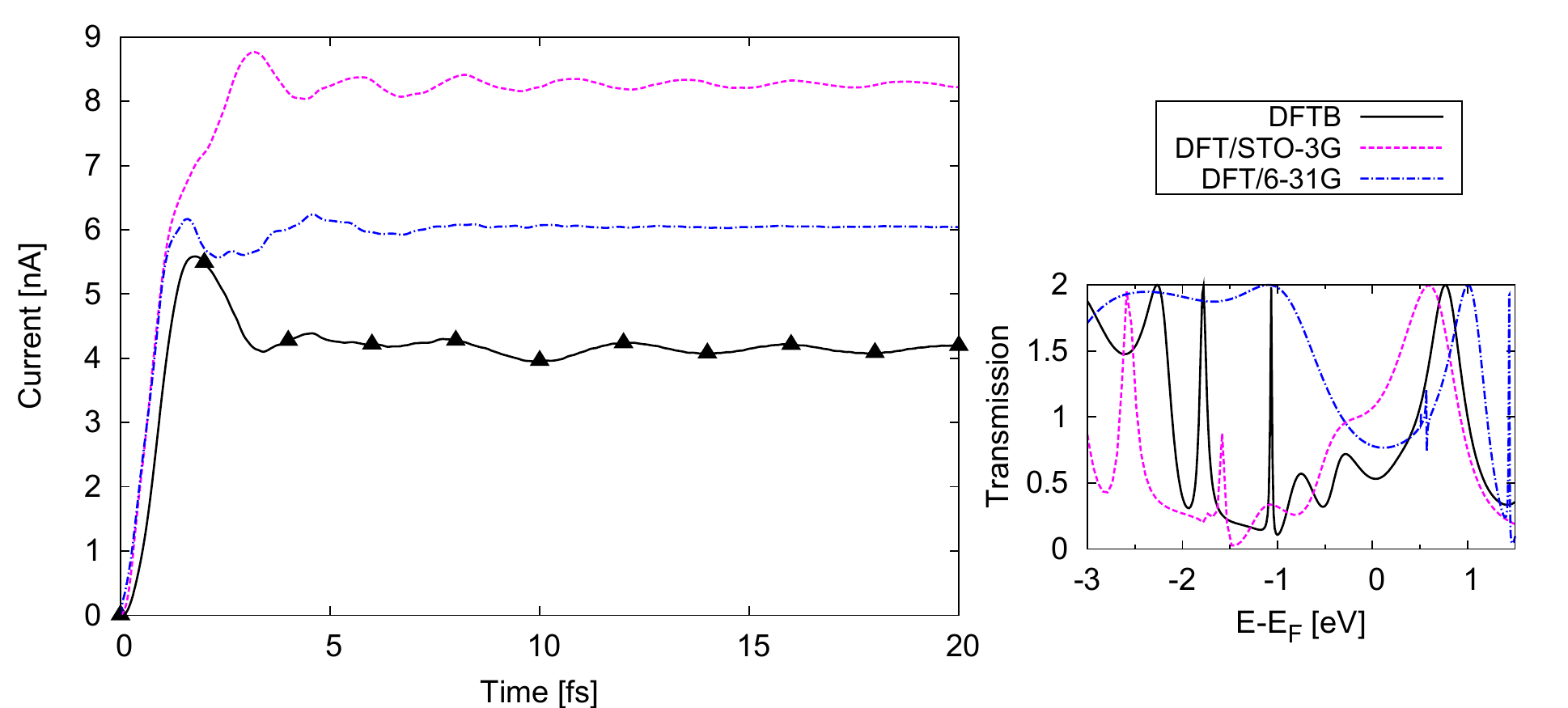}
\caption{Left figure: Time dependent currents for the carbon chain
  device of Fig.~\ref{LDR}b as given
  by TD-DFTB and TDDFT using the LDA exchange-correlation functional
  with the two basis sets STO-3G and 6-31G. Simulation parameters:
  $V_L$ = 0 V, $V_R^0$ = 0.1 mV, T = 1 fs. The solid line for the
  TD-DFTB method is the current at the left lead-device interface,
  while the filled triangles give the current at the right lead-device
  interface. Right figure: Transmission functions T(E) at $V$ = 0 V obtained from the
  static DFT-NEGF approach at the respective level of theory. }
\label{linrespo}
\end{center}
\end{figure}

In this section we benchmark the approximate TD-DFTB
scheme against results obtained with the
{\tt LODESTAR} code, that implements the reduced density matrix
propagation method of Sec.~\ref{secrdm} at the full TDDFT
level. Throughout the paper the Perdew-Burke-Ernzerhof
exchange-correlation functional is employed for TD-DFTB and the
TDDFT calculations are performed in the adiabatic LDA. Two different
Gaussian-type basis sets are used in the latter: the STO-3G set is roughly
of the same size as the minimal DFTB basis, while the more accurate
6-31G set is a split-valence basis of double-$\zeta$ quality.

We will first discuss the regime of linear response with small bias. Results for the temporal bias profile of Eq.~(\ref{btd}) with $T=$ 1 fs
and $V_R^0 =$ 0.1 mV are given in Fig.~\ref{linrespo}. Currents can be
evaluated at either the left or right lead-device interface. In the
initial phase these current may in principle differ in magnitude,
indicating a transient charging and decharging of the device. In the
steady state both need to be equal, which provides a stringent
test for the numerical accuracy of the implementation. 

Comparison of the current traces shows that a steady state is reached in roughly 5 fs, although especially in the TDDFT/STO-3G and TD-DFTB methods small amplitude oscillations are observed which fully die out only after several tenths of fs. The fact that the current does not follow the bias instantaneously was related to a kinetic inductance in Ref.~\cite{ke2010time}, which originates from the inertia or effective mass of the carriers in the leads.
Focussing now on the long time limit and taking the
6-31G results as reference, the steady state currents of TD-DFTB and
TDDFT/STO-3G show a deviation of roughly 30 \%. The conductance of the
latter is with $G=$1.1 $G_0$ ($G_0=$ 77.48 $\mu$S) in good agreement
with the single-$\zeta$ results of Ke {\em et.al.~} \cite{Ke2004}.

One drawback of time-dependent transport simulations are the limited
options to analyze the outcome. In contrast, static DFT-NEGF
simulations are usually discussed in the framework of Landauer
transport theory which features a bias and energy dependent transmission
function T(E,V) as key object \cite{Cuevas2010}. The  transmission
provides detailed information on the molecular resonances, as well as
on the device-lead coupling and the transparency of individual
transport channels. {\em A priori}, it is not obvious that TDDFT
simulations follow the Landauer picture and recent investigations show
that this is indeed not the case
\cite{Evers2004,Sai2005,Vignale2009}. Using local or semi-local
exchange-correlation functionals, however, steady state
TDDFT and static DFT-NEGF currents agree with high precision
\cite{Yam2011}. For the present system we find for example a DFTB-NEGF
current of 4.14 nA compared to a TD-DFTB value of 4.19 nA. In order to
facilitate a deeper analysis, Fig.~\ref{linrespo} also contains the
transmission functions obtained at the different levels of theory from
static DFT(B)-NEGF simulations in the WBA. In the linear response
regime, the current is governed by the transmission T($E_F$) at the
Fermi energy, which shows the same order (DFT/STO-3G $>$ DFT/6-31G $>$
DFTB) as found for the steady state current. The high transmission of
the STO-3G  mainly stems from the tail of a transmission channel at 0.5
eV above $E_F$, which is found at higher energies for the other two
methods. To the 6-31G transmission at $E_F$ contributes also a broad
feature centered around
-1.0 eV, which splits into narrower peaks at the DFTB level. This fact indicates
a smaller imaginary part of the self-energy for this channel, which
might also explain the slower decay of oscillations in the transient
TD-DFTB current (Fig.~\ref{linrespo} left).

\begin{figure}
\begin{center}
\includegraphics[angle=0,scale=0.7]{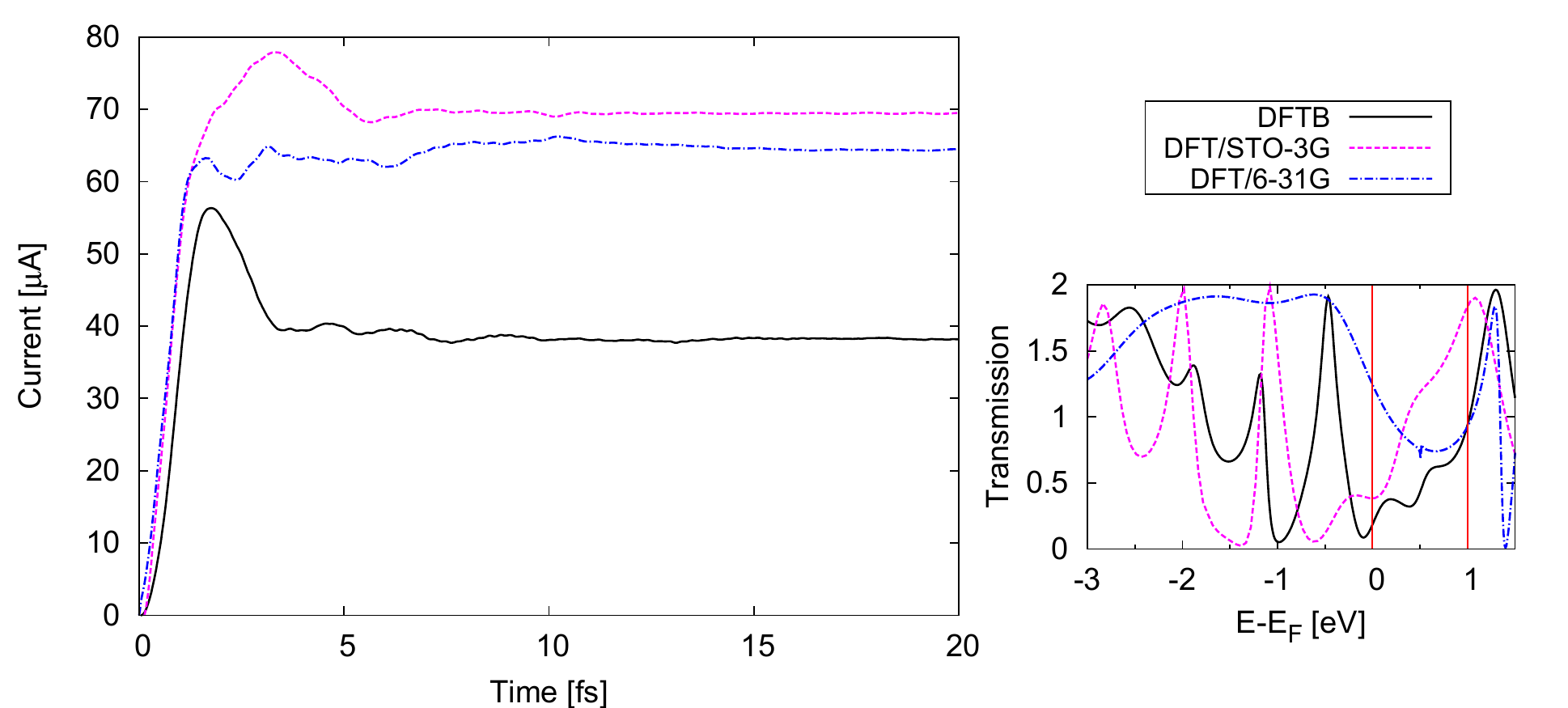}
\caption{Left figure: Time dependent currents for the carbon chain
  device. Simulation parameters:
  $V_L$ = 0 V, $V_R^0$ = 1.0 V, T = 1 fs. Right figure: Transmission
  functions T(E) at $V$ = 1 V. The vertical lines indicate the bias
  window.}
\label{c1v}
\end{center}
\end{figure}
Turning now to current traces at a higher bias value of $V =$ 1 V in Fig.~\ref{c1v}, we
find the two TDDFT results to be close, although this is somewhat
coincidental. At finite bias, the current in the Landauer picture is
obtained by integrating the transmission over the bias window from 0 V
to 1 V as indicated in the figure. For the STO-3G basis the resonance
above $E_F$ strongly contributes, whereas only a fraction of this
channel resides in the bias window for DFTB and DFT in the more
accurate 6-31G basis. DFTB resonances that are located in
equilibrium at -0.25 V and -0.75 V
with respect to $E_F$  shift into the bias window, but have a small
transmission leading to the lower value of the current observed.

The sizable variation of the transmission characteristics among the
different methods is also seen for the next device we
studied. Fig.~\ref{LDR}c shows the structure of 1,4-benzenediol
inbetween Al wires of the same kind as in the previous simulations. The
DFTB optimized molecule was placed symmetrically into the junction so
that the terminating oxygen atoms are positioned on the
hollow sites of the Al surfaces at a distance of 1.22 $\AA$. This
geometry is also used in the following TDDFT calculations.
\begin{figure}
\begin{center}
\includegraphics[angle=0,scale=0.7]{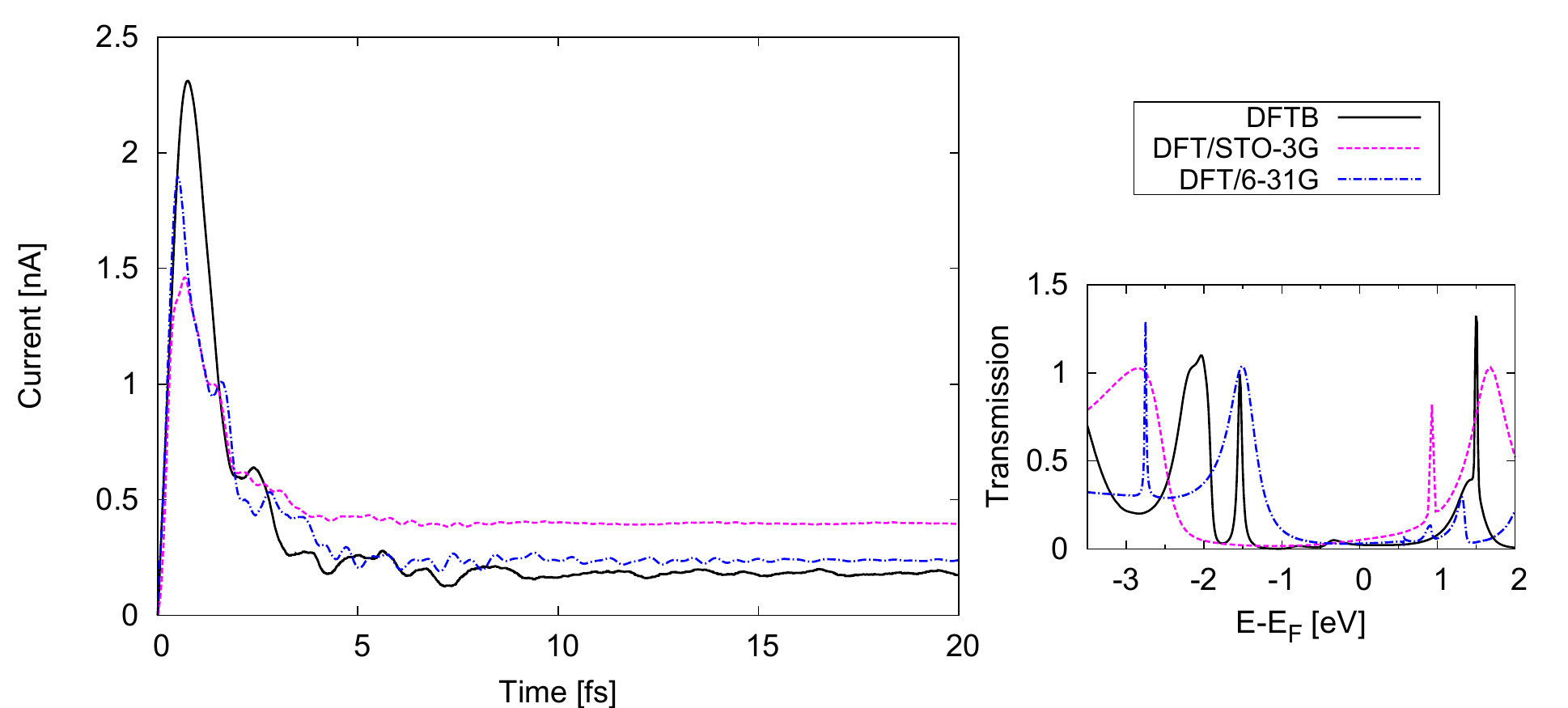}
\caption{Left figure: Time dependent currents through 1,4-benzenediol
  (see Fig.~\ref{LDR}c). Simulation parameters:
  $V_L$ = 0 V, $V_R^0$ = 0.1 mV, T = 1 fs. Right figure: Transmission
  functions T(E) at $V$ = 0 V.}
\label{benz}
\end{center}
\end{figure}

Focusing first on the left part of Fig.~\ref{benz}, one observes a
strong initial overshoot of the current in all methods, before it
reaches a constant value after roughly 5 fs. Again, the DFTB current
shows longer lasting oscillations than TDDFT/6-31G, but the limiting
values agree. Also in the transmission the DFTB results are
closer to TDDFT in the larger basis. T(E) features a transmission gap
of roughly 2.5 - 3
eV which is smaller than the corresponding gap for the isolated
molecule (C$_6$O$_2$H$_6$), for which we find a value of roughly 3.8
eV in all methods. The highest occupied molecular orbital (HOMO) is
fully transparent at the DFTB and DFTB/6-31G level and located at
-1.5 eV in Fig.~\ref{benz}, with a considerably reduced width in the
former method. Inspection of the local density of states (LDOS, not shown
here) reveals that also the DFT/STO-3G scheme has a state
 at -1.9 eV, which carries however no current and does not
appear in the transmission. The precise location of the lowest
unoccupied molecular orbital (LUMO) is more difficult to access. The LDOS
exhibits peaks at approximately 1.5 eV (DFTB), 0.9 eV and 2.0 eV (DFT/STO-3G) and  0.5 eV, 0.9
eV and 1.5 eV  (DFT/6-31G), respectively. Judging from the LDOS alone
it is difficult to discriminate between molecular states and metal induced
gap states that appear due to the presence of the additional Al layers in the extended
molecule. The latter are usually localized and do not significantly
contribute to the transmission. Taking this into account, we therefore tentatively assign the
LUMO state to the resonance at $\approx$ 1.5 eV for DFTB and DFT/6-31G
as well as $\approx$ 2.0 eV for DFT/STO-3G. These observations fit into the general trend  observed for gas
phase molecules: The DFTB method overestimates HOMO-LUMO gaps in
comparison with DFT calculations with converged basis sets, but to a
much lesser extent than DFT simulations with a basis of single-$\zeta$
quality \cite{Niehaus2001a}. The HOMO-LUMO gap has a strong impact on the transport characteristics of a device\footnote{More precisely, the equilibrium conductance is most strongly influenced by the level closest to $E_F$. Since $E_F$ is often located in the middle of the HOMO-LUMO gap, the size of the latter is also determining the transport gap.}  and it is known that DFT simulations overestimate conductances. This is due to the fact that the DFT HOMO-LUMO gap is significantly smaller than the true quasiparticle gap, given by the difference of ionization potential and electron affinity. In this regard, one would expect that the currents obtained from DFTB are closer to experimental values. This is certainly another case of achieving the right result for the wrong reason, but this point could be of interest for investigations targeting pragmatic solutions.

Returning to Fig.~\ref{benz}, it is obvious that $T(E_F)$ and
therefore also the current is very sensitive to the precise peak
position and broadening of the molecular states. In fact, it is not
uncommon that transport simulations using slightly different levels of
theory or even different implementations of the same level of theory
differ by one order of magnitude in the predicted currents
\cite{Lindsay2007}. In this respect, the TD-DFTB scheme falls well into these
general error bars.

At the end of this section, we shortly report the required CPU time
for the simulations presented. On a single core of an {\tt Intel Xeon
  X5560@2.80GHz} multi-core processor, 100 time steps for the carbon
chain device took roughly 120 CPU minutes (TDDFT/6-31G), 43 minutes
(TDDFT/STO-3G) and 8 minutes (TD-DFTB), respectively. For the
1,4-benzenediol device, the simulations took 139 minutes (TDDFT/6-31G),
48 minutes (TDDFT/STO-3G) and 7 minutes (TD-DFTB), respectively.

\section{Summary and outlook}
\label{summ}
In this article, we presented theory, implementation and validation of an approximate TDDFT method for open systems out of equilibrium. The temporal profile of the TD-DFTB current traces was found to be qualitatively similar to {\em first principles} simulations.  A rapid switch leads to an overshoot of the current which settles into a steady state after only a few fs. As in earlier TDDFT simulations \cite{Yam2011}, a steady state is always reached and does not depend on the history of the applied bias. Notwithstanding, we showed that permanent current oscillations are valid solutions of our employed equations of motion. The absolute value of the steady state currents was then discussed in the language and framework of the conventional static DFT-NEGF formalism. We found that TD-DFTB and TDDFT can differ significantly from each other, but this difference is not larger than the one between full TDDFT calculations using different basis sets. With respect to computational efficiency, the TD-DFTB approach is roughly one order of magnitude faster than a {\em first principles} implementation in the same formalism.  

A limitation of the presented approach is related to applications which require a large bias voltage of several V. In this case also molecular resonances far away from the frontier orbitals are moving into the bias window which, especially for the unoccupied states, are insufficiently described at a minimal basis set level. As mentioned earlier, it is entirely possible to include higher angular momentum states in the DFTB basis, although the two-center approximation for the Hamiltonian matrix elements becomes less accurate in this case. A larger basis does therefore not necessarily lead to a better result. 

The main advantage of the TD-DFTB scheme is its computational efficiency. This allows for important extensions which are not easily realizable in a {\em first principles} framework. To this class belongs for example the implementation of the hierarchical equation of motion approach \cite{Zheng2010}, which goes beyond the wide band approximation for the leads. Conceivable is also the consideration of quasiparticle corrections to the transport gap. For the static case, such GW calculations have already been performed within the DFTB framework \cite{Niehaus2005a,GagliardiPNFD2007}. Having applications to molecular photoswitches in mind, another important extension would be the realization of molecular dynamics simulations under current flow involving time-dependent external fields. Efforts in this direction are currently under way.
 
 \section*{Acknowledgement}
The authors would like to thank the German Science Foundation (SPP 1243 {\em Quantum transport at the molecular scale}) for financial support. 

\bibliographystyle{elsarticle-num}
\bibliography{../../Combined}
\end{document}